# A NEXAFS STUDY OF NITRIC OXIDE LAYERS ADSORBED FROM A NITRITE SOLUTION ONTO A Pt(111) SURFACE

M. Pedio[1], E. Casero[5], S. Nannarone[1,2], A. Giglia[1], N. Mahne[1], K. Hayakawa[3,8], M. Benfatto[3], K. Hatada[3], R. Felici[4], J.I. Cerdá[7], C. Alonso[6], J. A. Martin-Gago[7]

[1]*TASC National Laboratoy CNR-INFM, Trieste, Italy*
[2]*Università di Modena e Reggio Emilia, Dip. Ingegneria dei Materiali e dell'Ambiente, Modena, Italy*
[3]*Laboratori Nazionali di Frascati, LNF- INFN, Frascati, Italy*
[4]*ESRF, Grenoble, France*
[5]*Dpto. Química Analítica y Análisis Instrumental. Universidad Autónoma de Madrid. 28049 Madrid. Spain*
[6]*Dpto. Química-Física Aplicada. Universidad Autónoma de Madrid. 28049 Madrid. Spain*
[7]*Instituto de Ciencia de Materiales de Madrid-CSIC. Cantoblanco 28049-Madrid. Spain*
[8]*Museo Storico della Fisica e Centro Studi e Ricerche ``Enrico Fermi'', Via Panisperna 89A, 00184 Roma, Italy*

**ABSTRACT**

NO molecules adsorbed on a Pt(111) surface from dipping in an acidic nitrite solution are studied by near edge X-ray absorption fine structure spectroscopy (NEXAFS), X-ray photoelectron spectroscopy (XPS), low energy electron diffraction (LEED) and scanning tunnelling microscopy (STM) techniques. LEED patterns and STM images show that no long range ordered structures are formed after NO adsorption on a Pt(111) surface. Although the total NO coverage is very low, spectroscopic features in N K-edge and O K-edge absorption spectra have been singled out and related to the different species induced by this preparation method. From these measurements it is concluded that the NO molecule is adsorbed trough the N atom in an upright conformation. The maximum saturation coverage is about 0.3 monolayers, and although nitric oxide is the major component, nitrite and nitrogen species are slightly co-adsorbed on the surface. The results obtained from this study are compared with those previously reported in the literature for NO adsorbed on Pt(111) under UHV conditions.

1. **Introduction**

In the last few years a great interest has been paid to the use of surface science characterization techniques to gather insights into non-perfect systems. This is particularly true for structures prepared in aqueous conditions. Due to its extremely low cost, layers prepared in this environment are more suitable for technological applications than those prepared in ultra high vacuum (UHV) conditions. One of the main current concerns for surface science scientists is the understanding of the so-called *pressure gap* issue in the formation of surface structures. In this sense, it will be of a valuable interest to know whether the results obtained during the last 20 years in surface science can be extrapolated to 'real' systems. That is, to know whether there is a significant difference in the adsorption sites, coverage and long- and short-range ordering of a molecule adsorbed either from a liquid environment or under ultra high vacuum conditions.

We have addressed this issue by studying the adsorption of nitric oxide on a clean Pt(111) surface from an acidic nitrite solution. Subsequently, we have compared the results obtained from this study with those previously reported in the literature for NO adsorbed on Pt(111) in UHV conditions. NO/Pt(111) has been chosen as a model system because of its importance in the environmental chemistry. Nitric oxide is a strong pollutant and one of the major components of automotive exhaust gases. The removal of this harmful component is commonly achieved by the use of catalytic materials, such as platinum, which allows the catalytic conversion of the hazardous $NO_x$ species to innoxious reaction products. Therefore, insights into the relevant reaction mechanisms are crucial to understand the fundamental properties of real catalytic systems allwoing to optimize these processes.

Although, NO adsorption on Pt(111) surfaces has received great attention during the past years, the question of whether or not the results obtained in UHV can be extrapolated to real catalysis remains open. The studies of NO adsorption on well-defined platinum surfaces are usually carried out either from the gas phase in Ultra High Vacuum (UHV) environments or from a solution in an electrochemical cell. In the latter case, the adsorption is performed from a solution saturated with NO gas (or from acid solution of nitrite ions).

All recent works concerning NO adlayers formed in UHV conditions [i,ii,iii,iv,v,vi,vii,viii] conclude that NO adsorbs molecularly between 100 and 375 K, with the N atom bound to the Pt surface, and with a small tilt of the molecular axis with respect to the surface normal. Over this range of temperatures, many different phases and adsorption sites have been reported. However, it seems clear that the saturation coverage at RT is about 0.5 ML.

On the other hand, studies concerning adsorption of NO on Pt(111) from solution conclude that NO can adsorb molecularly either from a solution saturated with NO gas or from an acid solution of nitrite ions, leading to saturation coverages smaller than 0.5 ML. The structural characterization of the NO adlayers obtained from solution has been performed by many different techniques [ix]. We highlight the results obtained from two works based on in-situ STM images. Both studies investigate the structure of the NO adsorption on Pt(111) in an electrochemical cell and the further oxidation under potential control. [x,xi]. From the structural point of view, both studies have reported ordered adlattices at 0.7 V (reduced phase) vs. normal hydrogen electrode (NHE) exhibiting a (2 x 2) periodicity. However, for the oxidized phase (at 1.0 V vs. NHE) Z. Zang et al. [10] reports a (√13 x √13)R14º whereas K. Momoi et al.[11] reports a (2 x 2) structure. Interestingly, and in opposition to the results presented in such references, two other works [xii,xiii] performed *in* and *ex situ* showed that there is no long range ordered phase. From these studies, it is clear that controversy regarding the bonding geometry and configuration of nitric oxide on Pt(111) when the adsorption is performed in an electrochemical cell still remains open.

In this paper, we present a detailed NEXAFS characterization study of nitric oxide adlayers adsorbed from an acidic nitrite solution on a flame annealed Pt(111) surface. We compare the structural results concerning structure, adsorption site occupation and saturation coverage, derived from NEXAFS, LEED, XPS and STM measurements for our preparation method (NO adsorption from solution) with the results reported in the literature for UHV conditions. We show that NO is adsorbed molecularly, with a small saturation coverage (less than 0.3 ML) and coexisting with small amounts of $N_2$, CO and $NO_2$.

**2. Experimental and Theoretical**

*2.1. NO adlayers preparation on Pt(111)*

The Pt(111) single crystal preparation procedure consisted of annealing the crystal to about 1000ºC in a gas/oxygen flame for 3 min and then allowing it to cool for 60s in the vapor of deaerated supporting electrolyte before quenching in the same solution. Surface cleanness was determined by cyclic voltammetry in 0.1 M $HClO_4$. Voltammograms of the clean Pt(111) show the characteristic "butterfly" shape first described by Clavilier [xiv] (data not shown). The experimental set-up consisted of a three-compartment cell with provision for the addition and withdrawal of solutions under a positive pressure of pre-purified nitrogen at all times.

After recording the characteristic voltammogram of the clean Pt(111) in 0.1 M $HClO_4$, this solution was replaced by a $10^{-2}$ M $KNO_2$ + 0.1 M $HClO_4$ solution in a flow-through fashion inside the cell. NO adlayers were generated by holding the clean Pt(111) electrode at open circuit for 3 min in contact with this solution. The nitrite solutions were prepared from $KNO_2$ (97%) and perchloric acid (70%, 0.1M), both reagents Merck. All these solutions were prepared from water purified with a Millipore Milli-Q system (18MΩ).

*2.2 Experimental techniques*

NEXAFS and XPS measurements were performed at BEAR beamline of the ELETTRA synchrotron radiation facility in Trieste, Italy[xv]. Photoemission from the N 1s, O 1s and Pt 4f core levels were measured by using a double pass cylindrical mirror analyser (CMA). The vertical exit slit of the beam line was 100 μm corresponding to a photon energy resolution of 500 meV at the N K-edge and 730 meV at O K-edge. The NEXAFS spectra at O and N K-edges are normalized to the incident photon flux obtained by measuring simultaneously the current drained by a tungsten grid inserted in the beam path. The tungsten grid was previously heated at high temperature to remove contaminants. The NEXAFS spectra have been measured in Auger yield mode on the NO/Pt(111) sample and normalized to the cleaned surface measured as reference. The CMA was tuned at the N and O LVV Auger lines.

STM images were recorded on a commercial RT-STM. Images are presented in topographic mode. The base pressure in the ultra high vacuum (UHV)-chamber during measurements was better than $10^{-9}$ mbar. During transport from the electrochemical cell to the UHV chamber the samples were in a glass cell under nitrogen atmosphere while immersed in the corresponding $10^{-2}$ M $KNO_2$+0.1 M $HClO_4$ solution. By transferring the samples to the UHV chamber, they were briefly exposed ($\approx$ 5-10 sec) to the atmosphere. Direct contact with the air was, however, avoided by leaving a drop of solution on the electrode surface. The remaining solution was removed during pumping in the vacuum chamber.

*2.3 Theoretical details*

The NEXAFS spectra have been simulated by means of the MXAN code [xvi]. The X-ray photoabsorption cross section is calculated using the general Full Multiple Scattering scheme within the muffin tin (MT) approximation. For the exchange and correlation part of the potential we used phenomenological potentials: energy-independent Xa potential and the complex Hedin-Lundqvist (HL) [xvii]. The coordinate structure was fixed while some potential parameters were fit, such as the Fermi energy and the values of the interstitial potential and the overlap between contiguous atoms. Details will be provided in a forthcoming publication.

Additionally, we have also performed first principles total energy calculations under the Density Functional Theory (DFT) framework to study the NO adsorption on the Pt(111) surface. All calculations were performed with the SIESTA code [siesta] employing the Local Density Approximation (LDA)[lda]. To model the surface system, we considered a three layer thick Pt slab oriented along the (111) plus a NO molecule adsorbed at the top layer in a p(3x3) cell. A localized double zeta polarized (DZP) basis set was used to describe the valence electrons while the core electrons were replaced by Troullier-Martins pseusopotentials[pseudo]. All atoms in the slab were allowed to relax except the bottom Pt layer, which coordinates were fixed to the bulk Pt ones.

## 3. Results and discussion

### *3.1 Structure of the adlayer*

LEED patterns (data not shown) recorded after NO adsorption show the typical (1x1) surface symmetry of Pt(111). Just a small increase of the background after dosing with respect to the clean surface is observed. This is a clear indication that the saturation coverage is small, and no ordered super-structures are formed by this simple deposition procedure, in agreement to a previous grazing X-Ray diffraction study.[12]

To further confirm this structural picture of the NO adsorption on a Pt(111) surface, we have recorded STM topographic images of both a Pt(111) clean surface and a sample prepared in the conditions previously described (Fig.1). Pt(111) clean surface image (data not shown) shows the typical surface morphology, consisting of flat terraces of about 100 nm width separated by single height steps with an average height of about 0.2 nm. Figure 1 shows a detail recorded on a terrace of the NO dosed surface. Small ordered molecular patches are appreciated without continuity (see for instance central part of the image). We could not found long-range ordered structures, even at high adsorption times. This result is in good agreement with that reported in refs 12-13, where no surface diffraction peaks were observed. The small patches observed on the surface could correspond to molecular islands exhibiting small locally ordered structures. The saturation coverage, inferred from both, LEED and STM, has therefore to be very low, of the order of few tenths of a monolayer.

### *3.2 N and O K-edge NEXAFS measurements*

The use of the NEXAFS technique for studying this kind of systems presents two main drawbacks. First, the small saturation molecular coverage (see previous section) makes the detection of the adsorption signal intricate and special care should be paid to the normalization process. As a consequence, spectra present a low count-rate, which make difficult the interpretation of the features. Secondly, due to the ex-situ characterization some contamination of air-contained species has an effect on the surface, rendering difficult a direct assignation of components to single molecular species. This is particularly true for oxygen and carbon edges.

Fig. 2A shows NEXAFS spectra -recorded in Auger-yield mode- of the N K-edge of the NO adlayer deposited from solution, with the polarization vector (E) of the impinging light in plane (normal incidence) and almost orthogonal (grazing incidence, angle of incidence 15°) to the surface. Fig. 2B shows the NEXAFS spectra of the O K-edge in Auger–yield mode in the same measurement geometries. Both spectra are rich in structures, indicating the co-presence of different N and O containing species. Table 1 shows the energy position of the features measured in Fig. 2A and Fig. 2B. Assignments of these features will be performed (see discussion below) following previously published results, which are summarized in Table 2. This table reports the published values for the π and σ resonances measured by NEXAFS of molecular NO, $NO_2$, $O_2$, $N_2$ and CO in gas phase and adsorbed onto Pt(111) surfaces in UHV conditions.

In N K-edge the π* resonance region shows two clearly resolved components at energies of 400.5 eV and 401.2 eV (labeled A and B in Fig. 2A). These components have been previously identified for the UHV deposition at 230 K as π* resonances of the NO molecule.[6] This splitting has been reported for other molecules. [6, xviii,xix,xx].

The structure labelled D in fig. 2A is much broader and it emerges in the energy region related to σ* molecular transitions. The position of this feature is normally most affected by the interaction with the chemical environment of the element, such as intramolecular bond changes and substrate interaction, as discussed by J. Stöhr[18] (rule of thumb) and reproduced by multiple scattering calculations (MSC). [16,17,18,xxi]. The resonance D lays at about 12 eV above the π* resonance region and therefore it can tentatively be assigned to the σ* NO shape resonance (see table 2). The D peak intensity results maximized at grazing incidence (whereas the A and B peaks are reduced), i.e. in a geometry with the E vector out of the surface plane and close to the surface normal, confirming the σ character of this feature. These observations confirm that the molecular orientation of the NO species correspond to the NO bonding in an up-right position with respect to the surface[18].

The features labelled as C (≈ 404 eV) and E (418 eV) in Fig. 2A cannot be directly ascribed to any NO peak. The NEXAFS spectra of $NO_2$ deposited in UHV on Ni(100) [xxii] show two π resonances centered at about 402.5 eV and a σ resonance at about 412.7 eV. The first peak could therefore correspond to the peak labelled C, whereas the tail of the σ* NO resonance (peak D) would hide the former peak. The peak labelled E could be assigned to the σ $N_2$ resonance. We will show in the next section that $N_2$ and $NO_2$ species coexist on the surface at smaller concentrations. Finally, the small pre-peak labelled 'At' at about 1 eV below the feature labelled A has not been previously reported for NO adsorption in UHV conditions. This peak can be ascribed to atomic N, following table 2.

Fig. 2B shows the O K-edge spectra. Although this peak could include some contribution from air contamination, the feature A at 531.5 eV can be assigned to π* NO resonance, in agreement to the NEXAFS measured by F. Esch [xxiii]. The feature B at 532.2 eV is ascribed to π* $NO_2$ according to table 2. The broad features C and D (544.0 eV and 547.8 eV, respectively) appear in the σ energy region of bi-atomic molecules[18] and can be assigned to σ NO and σ $NO_2$ resonances, respectively. [6] However, this assignation is not unique. Following table 2, features B and D could correspond to the π and σ signal from CO on Pt(111). The dependence of the NEXAFS intensity with the light-incidence angle is in agreement with the previous assignment of an upright molecular geometry.

*3.3 XPS measurements*

The presence of NO as well as a fraction of $NO_2$ and other species has been further verified by XPS measurements. Fig. 3A shows the photoemission N 1s level of the NO adlayer taken at a photon energy of 630 eV. The spectrum is wider with respect to the same measurement performed after dosing in-situ on UHV systems[5]. The spectrum shows the main peak at a binding energy of 400.5 eV and two distinct shoulders. A deconvolution of the peak shape in curves-components indicates the presence of three different components at 398.3, 400.5 and 402.4 eV, which could be assigned to $N_2$, NO and $NO_2$, respectively, in agreement with the results discussed in ref. 12 and the previous NEXAFS assignation. The relative intensities are 8% for the peak at 398.3 eV ($N_2$), 74% for that at 400.5 eV (NO) and 17% for the peak at 402.4 eV ($NO_2$), which could roughly correspond to the concentration of those species on the surface.

Fig 3B shows the O 1s level of the NO adlayer taken at a photon energy of 630 eV. The deconvolution of the spectrum shows also three components at about 530 eV, 531 eV and 533 eV. For comparison, the reported values for atomic O, NO, $NO_2$ and CO are also shown in the figure as vertical lines. Also this peak shows much broader widths with respect to the UHV prepared systems. These components can be related to the presence of NO and a small amount of atomic O [5 and refs therein], $NO_2$ [xxiv] and CO [xxv]. Due to the superposition of the lines related to NO and $NO_2$ (≈530 eV) and NO and CO (≈ 531 eV) an estimation of the fraction of the different O containing species is not feasible from the analysis of the O 1s core-level. The origin of CO and $N_2/O_2$ could be related to both contaminations from the flame annealing preparation method and the transport trough air.

*3.4 Multiple scattering and DFT ab-initio calculations*

We have applied multiple scattering calculations (MSC) to reproduce the spectroscopic features of the NO molecularly adsorbed on Pt(111) in UHV. The problem comes out from the presence of different molecules on the surface, which leads to many free parameters to derive accurate structural information. Thus, the determination of the orientation of the molecular axe from an analysis of the polarization dependence become much difficult due to both, the coexistence of different species on the surface and the low signal intensity, which makes the spectral line-shape quite complex. In any case, valuable information about the interatomic molecular distance can also be easily derived from the present data.

Multiple scattering simulations of the isolated NO and NO adsorbed on Pt(111) at atop, bridge and threefold hollow sites have been performed by MXAN code [16]. (see experimental details section). To learn about the influence of the surface on the molecular bonding distance, N-O, we have calculated the energy difference between π and σ resonances for several NO bonding distances, both for the N and O edge. The result is presented in Fig. 4. The interatomic distance of the NO molecule increase as the energy difference between the $\pi^*$ and $\sigma^*$ decrease. This relationship can be used to easily estimate the bonding intermolecular distance. The intramolecular N-O distance in gas phase is 1.15 Å[18], marked in the figure with a dashed vertical line. We have plotted in this graph the experimental values of the $\pi^*$-$\sigma^*$ energy difference for the NO species derived from Fig. 2A and 2B, for the N and O respectively as horizontal thick lines. Around those lines we have shadowed the experimental incertitude. The N experimental data suggests a bonding distance of about 1.22 ± 0.05 Å, which corresponds to a significant expansion of the molecular interatomic distance. For the O edge, we found a similar result, 1.20 ± 0.05 Å. The error bar comes mainly from the determination in the $\sigma^*$ value, which could be estimated to be about 1 eV. However, as can be seen in fig. 4, a large error bar value in the energy is reflected just a some hundreds on an Å in the interatomic distance.

The molecular expansion upon adsorption on the surface was further corroborated by the total energy calculations. Indeed, the optimized structures reveal an elongation of the N-O bond distance from 1.18 Å in the gas phase to 1.22 Å (1.21 Å) when it is adsorbed at the fcc (hcp) site. The calculated N-O bond distances are in excellent agreement with the MSC results. Therefore, we can conclude after the DFT and MS calculations for the XANES spectra that there is about 6% expansion in the N-O interatomic distance as a consequence of the NO adsorption onto Pt(111). Obviously, such expansion simply reflects the weakening of the NO bonding as a consequence of the N-Pt bonding. Therefore, although the XANES spectra are noise because of the low-coverage, the good agreement experiment-theory gives confidence in the experimental data.

**Conclusions**

The adsorption of NO on a flame annealed cleaned Pt(111) surface from dipping in an acid nitrite solution was studied by near edge X-ray absorption fine structure spectroscopy (NEXAFS), X-ray photoelectron spectroscopy (XPS), low energy electron diffraction (LEED) and scanning tunneling microscopy (STM) techniques. LEED patterns and STM images show that no long range ordered structures are formed after NO adsorption on a Pt(111) surface. NEXAFS and XPS spectra confirm that the main specie adsorbed in these conditions is nitric oxide. The NO molecule adsorbs molecularly on the surface trough the N atom in a defined orientation being the maximum saturation coverage about 0.2 monolayers. Besides the NO molecule, $NO_2$, and $N_2$ species are present on the surface in small amounts. These additional species do not show defined adsorption geometry, being their contribution to the spectra not dependent on the measurement geometry. Also some small traces of CO molecules are adsorbed on the surface, in an upright conformation. An expansion of the NO bonding distance of about 6% is inferred from the MS calculations and corroborated by DFT calculations.

**Acknowledgments:**

TASC technical service is kindly acknowledge for assistance. We are grateful to the Spanish agency CICYT for financial support trough the projects MAT2005-3866 and MAT2007-66719-C03-02 and the Consolider contract (NanoSELECT).

| N K-edge | | | O K-edge | | |
|---|---|---|---|---|---|
| Label | Energy | Resonance | Label | Energy | Resonance |
| A | 400.5 | π1 NO | A | 531.5 | π NO |
| B | 401.2 | π2 NO | B | 532.2 | π $NO_2$/ CO |
| C | 403.5 | π $NO_2$ | C | 544.0 | σ NO |
| D | 412.6 | σ NO | D | 547.8 | σ $NO_2$/ CO |
| E | 418.1 | σ $N_2$ | | | |
| At | 398.0 | Atomic N | | | |

Table 1. Energy position (eV) of the features measured in Fig. 2A and Fig. 2B and their corresponding assignments.

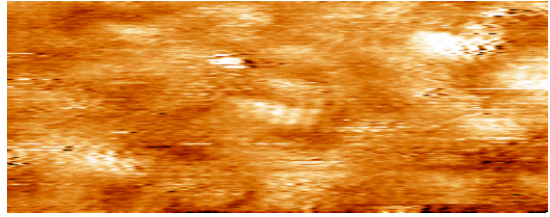

Fig.1. STM image (2.3nm witdh) in which some ordered patches could be seen in the central part of the image.

| System | Reference | Element | Incidence geometry | Energy (eV) $\pi$ resonance | Energy (eV) $\sigma$ resonance | $\Delta(E_\pi - E_\sigma)$ |
|---|---|---|---|---|---|---|
| **NO gas phase** | 18 | N | | 399.7 | 415 | 15.3 |
| | 18 | O | | 532.7 | 547 | 14.3 |
| **NO/Pt(111)** | 6 | N | NI (290K) GI (290K) | 401.1 401.1 | 412.5 | 11.4 |
| | | | NI (230K) GI (230K) | 400.7+401.4 401 | 412.5 | 11.5 |
| | | O | NI (290K) GI (290K) | 531 531+533 | 545.5 543 | 14.5 11.0 |
| | | | NI (230K) GI (230K) | 532 532.5 | 546 543 | 14.0 10.5 |
| **NO/Pt(111)** | 23 | O | (1 ML) | 532 | 543 | 11.0 |
| | | | NI (5ML) | 532 | 545.2 | 13.2 |
| | | | GI (5ML) | | | 11.0 |
| **NO₂ gas phase** | 24, 22 | N | | 402-404 | 418 | 15 |
| | | O | | 530-533 | 541-548 | 13 |
| **NO₂/Ni(100)** | 22 | N | | 402.5 | 412.7 | 10.2 |
| **O₂ gas phase** | 18 | O | | 531 | 541.5 | 10.5 |
| **O₂/Pt(111)** | 18 (physisorbed) | O | | 530.8 | 540.5 | 9.7 |
| | 18 (chemisorbed) | O | NI | 530.3 | 535.5 | 5.2 |
| | | | GI | 530.9 | | |
| **N₂ gas phase** | 18 | N | | 400 | 418 | 18 |
| **CO/Pt(111)** | xxvi | O | | 532.3 | 548.5 | 16.2 |

Table 2. Reported energy values of the $\pi*$ and $\sigma*$ resonances measured by NEXAFS of molecular NO, NO₂, O₂, N₂, and CO in gas phase and adsorbed onto Pt(111) or Ni (100) in UHV conditions.

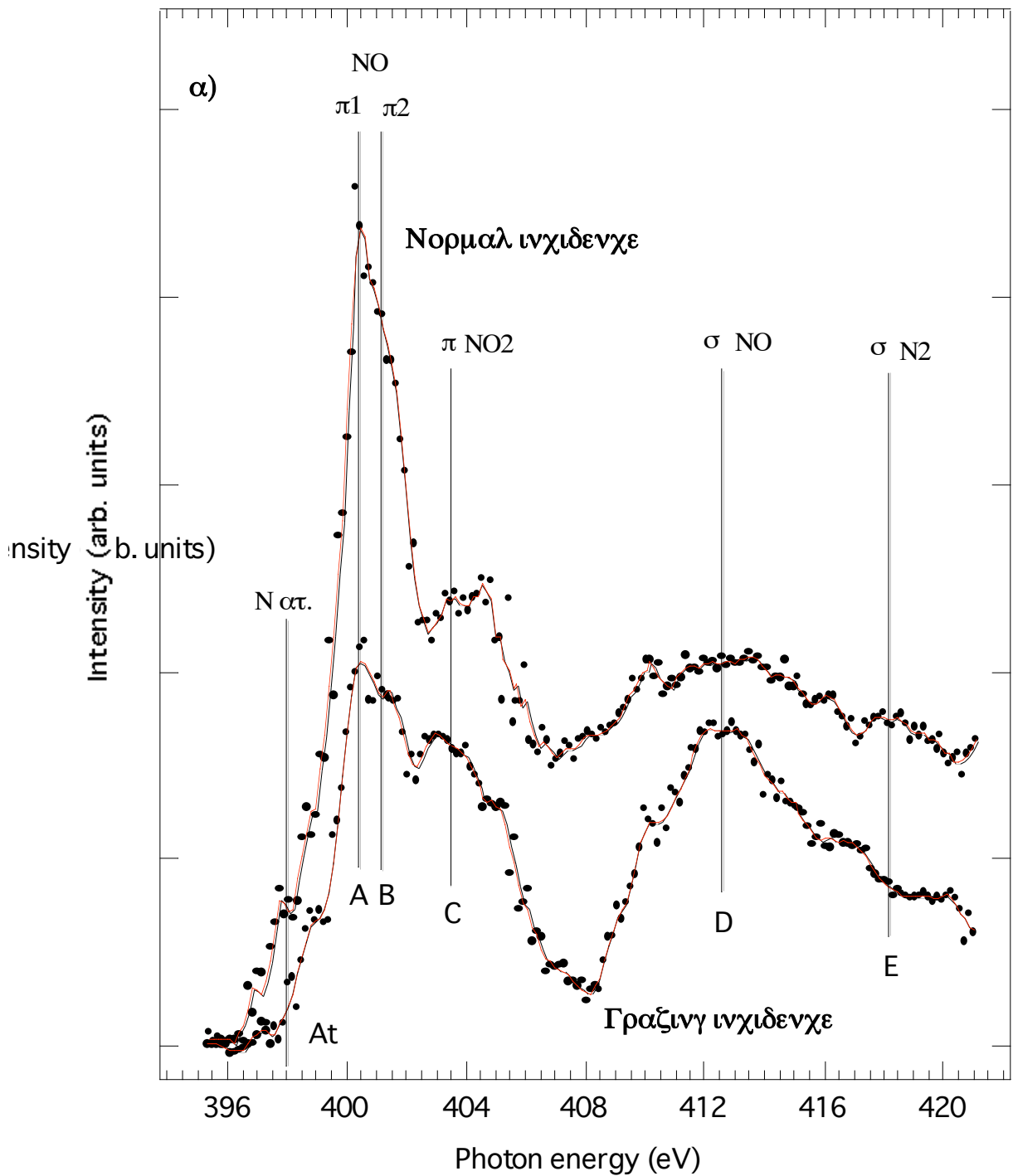

Fig. 2A. Absorption coefficient from Auger-yield of N K-edge of NO adsorbed on Pt(111) from solution taken at two polar angles: normal incidence and grazing incidence corresponding to the electric field in the surface plane and close to the surface normal, respectively. The spectral features are labelled At, A, B, C, D, and E. For the assignments see Table 1.

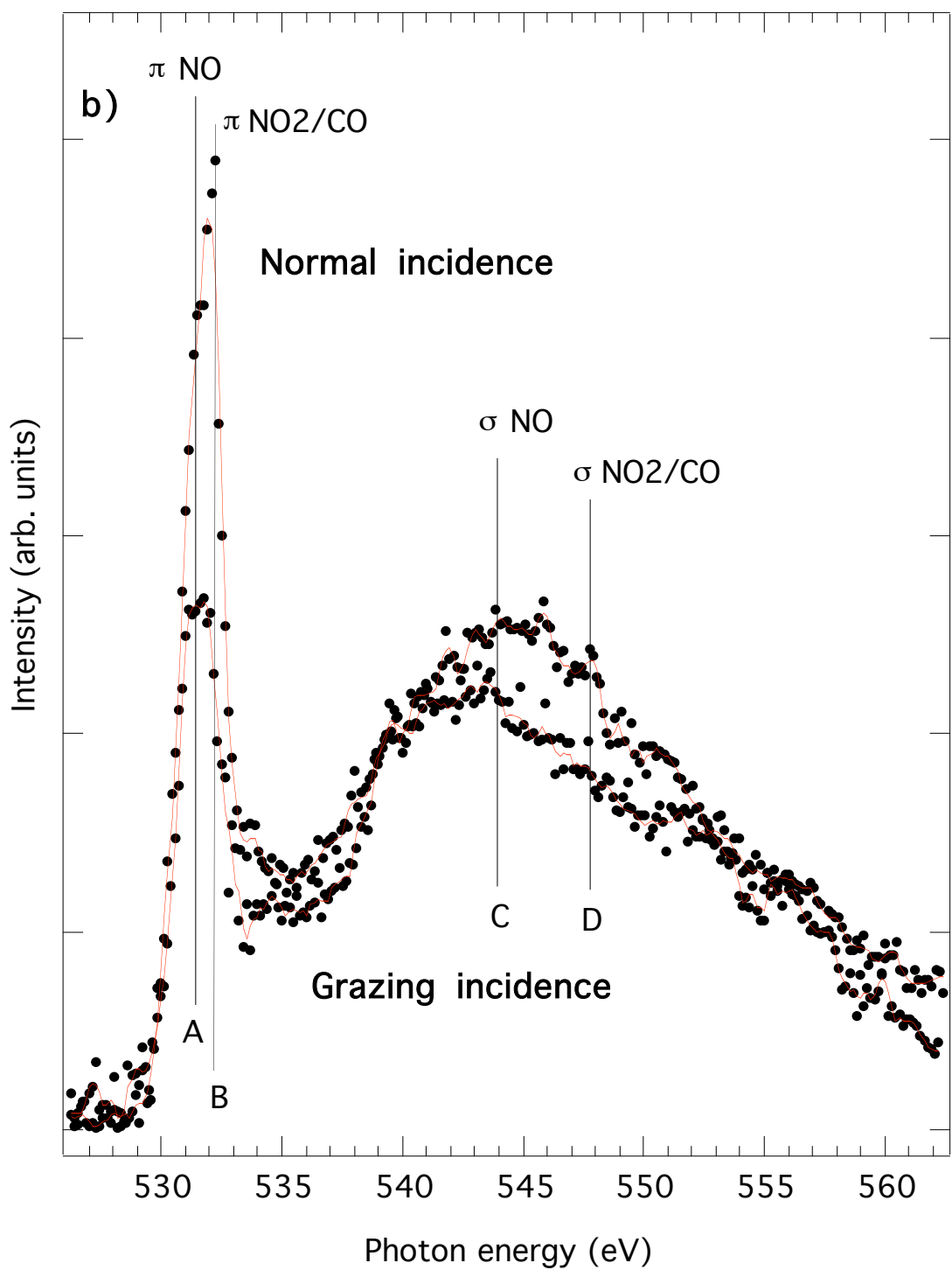

Fig. 2B. Absorption coefficient from Auger-yield of O K-edge of NO adsorbed on Pt(111) from solution taken at two polar angles: normal incidence and grazing incidence corresponding to the electric field in the surface plane and close to the surface normal, respectively. The spectral features are labelled A, B, C, and D. For the assignments see Table 1.

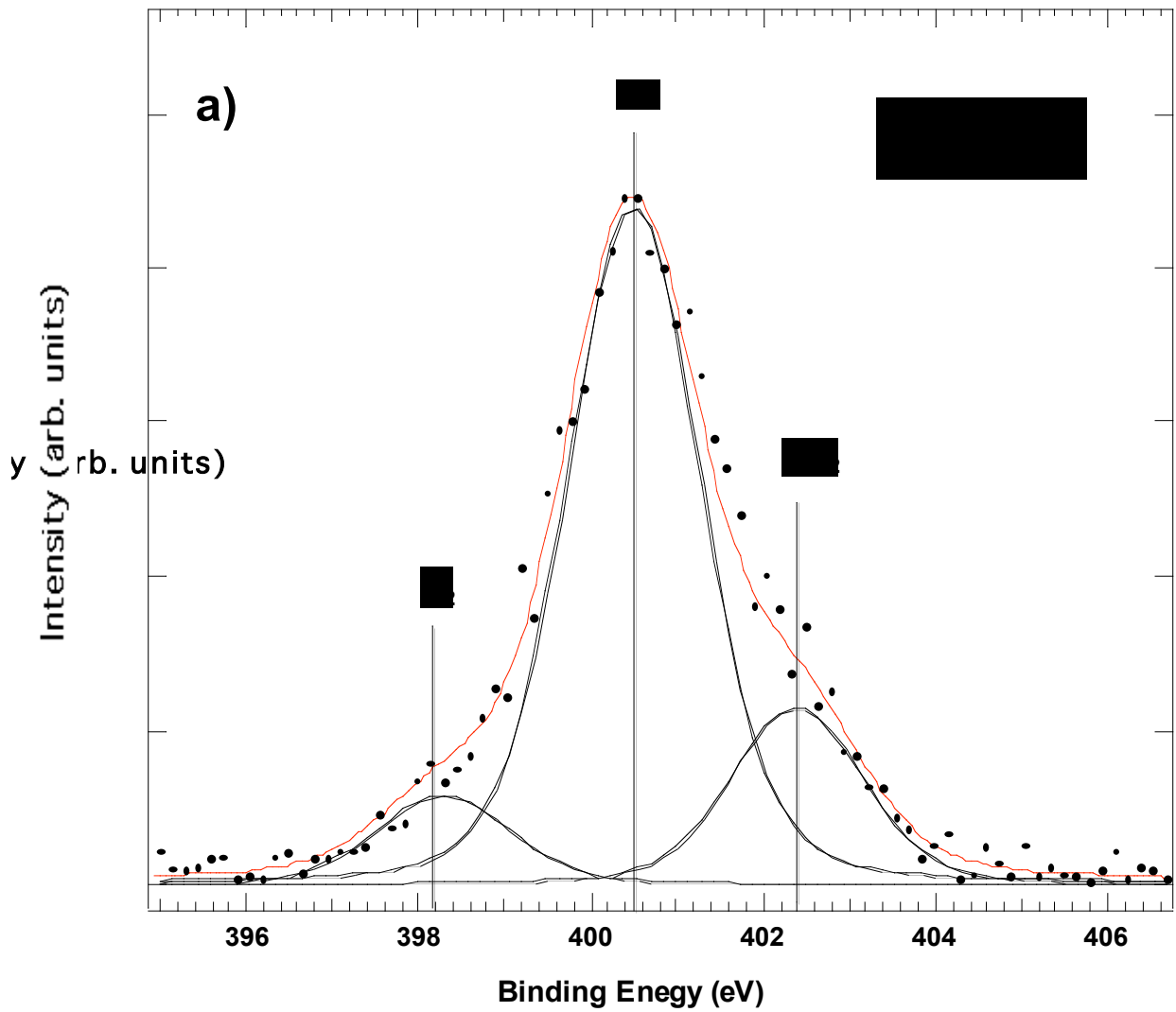

Figure 3A. X-Ray photoemission spectra of N 1s core level (after background subtraction) from NO adsorbed onto Pt(111). The assignments of the different N species are indicated (see also text).

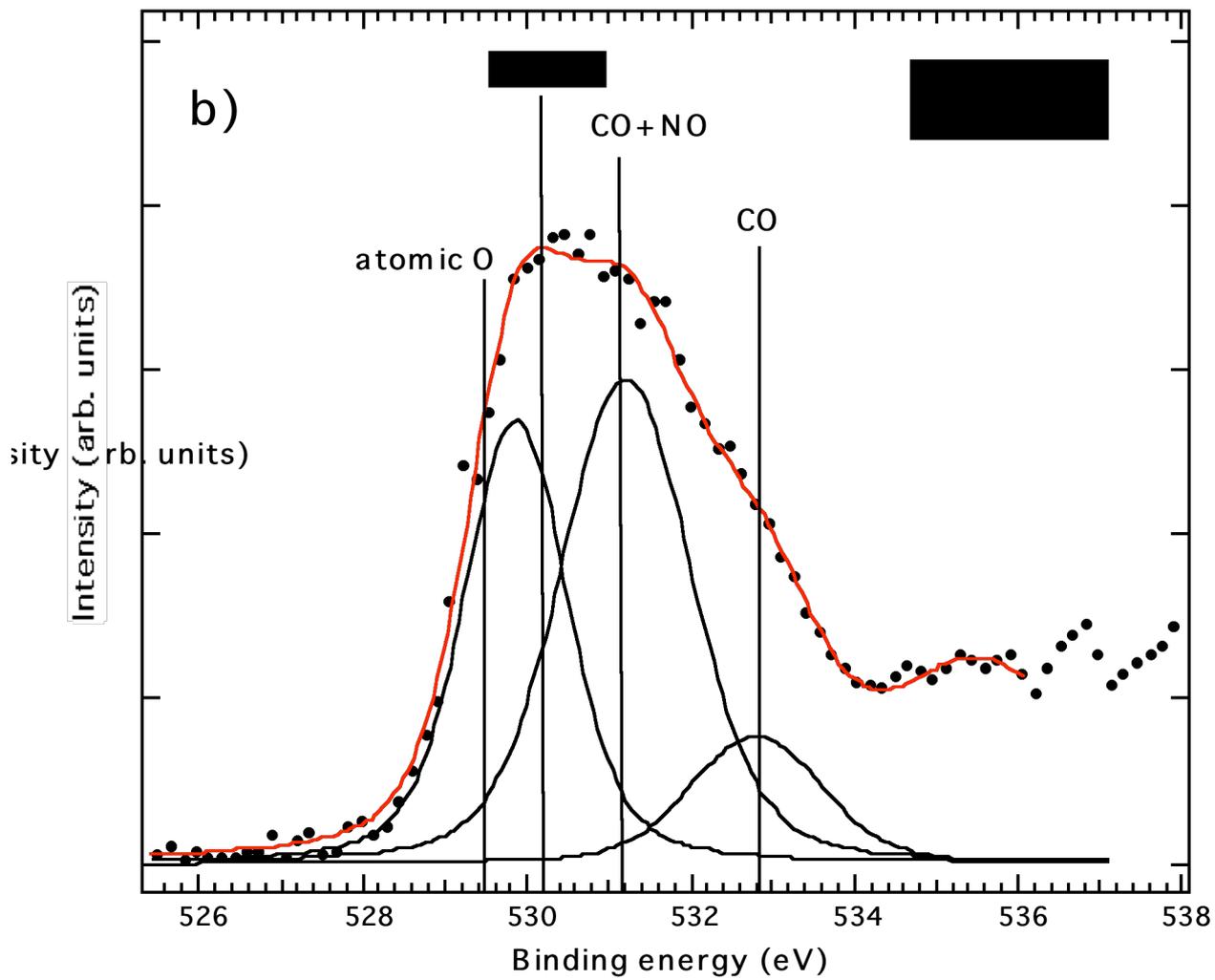

Figure 3B. X-Ray photoemission spectra of O 1s core level (after background subtraction) from NO adsorbed onto Pt(111). The assignments of the different O species are indicated (see also text).

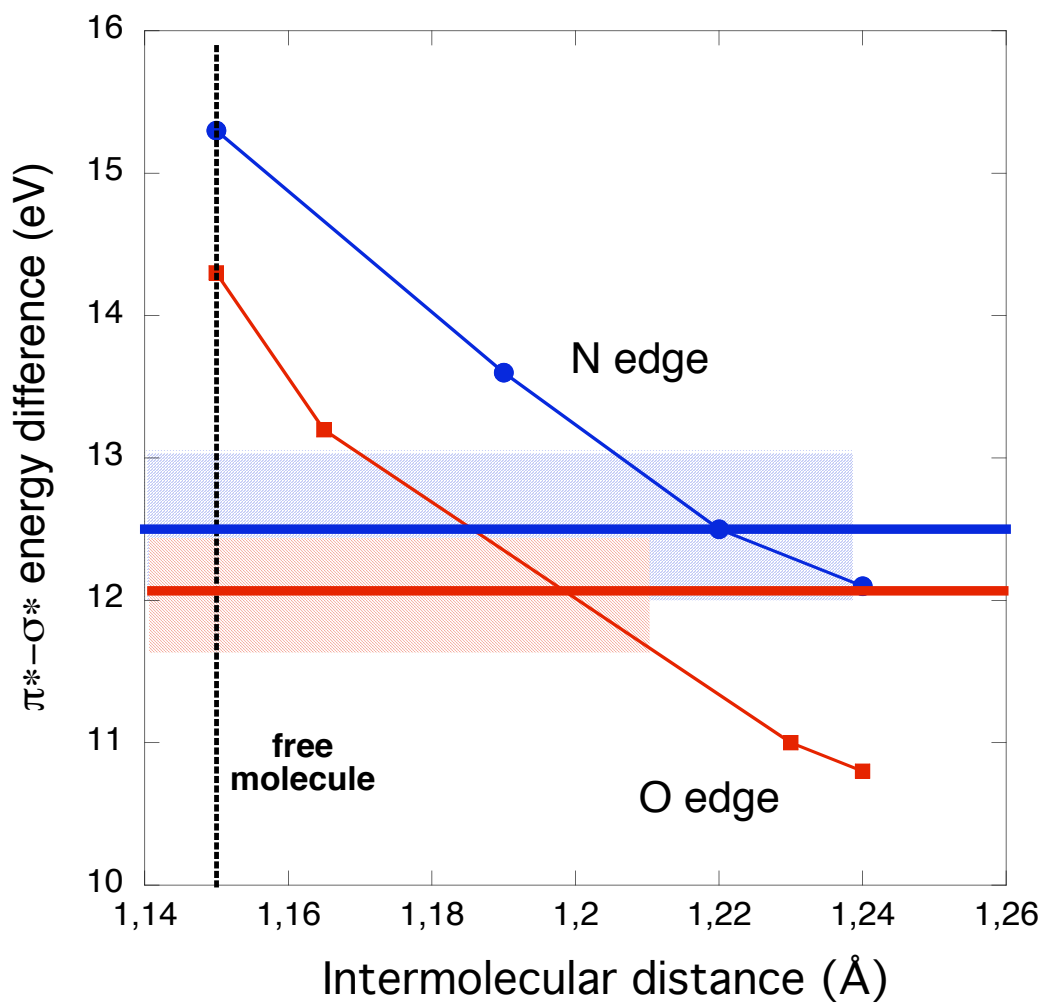

Fig. 4: The difference in energy between the π and σ resonances from MSC calculations as a function of the NO molecule bond length as estimated from absorption coefficient measurements of N 1s (open circles) and O 1s (filled squares). Parallel lines are experimental values taken from fig.2 and shadowed regions the experimental error bar.

# References

[i] H. Ibach, S. Lehwald, Surf. Sci. 76 (1978)

[ii] M. Kiskinova, G. Pirug and H. P. Bonzel, Surface Science 136 (1984) 285.

[iii] M. Matsumoto, N. Tatsumi, K. Fukutani, T. Okano, T. Yamada, K. Miyake, K. Hate, H. Shigekawa, J. Vac. Sci. Technol. A 17 (4) (1999) 1577

[iv] 1. M. Matsumoto, K. Fukutani, T. Okano, K. Miyake, H. Shigekawa, H. Kato, H. Okuyama, M. Kawai, Surf. Sci. 454–456 (2000) 101.

[v] F. Zhu, M. Kinne, T. Fuhrmann, R. Denecke, H.-P. Steinrück, Surface Science 529 (2003) 384–396

[vi] P. Zhu, T. Shimada, H. Kondoh, I. Nakai, M. Nagasaka, T. Ohta, Surface Science 565 (2004) 232–242

[vii] N. Materer, A. Barbieri, D. Gardin, U. Starke, J.D. Batteas, M.A. Van Hove and G.A. Somorjai, Surface Science 303 (1994) 319.

[viii] Q. Ge, D.A. King, Chem. Phys. Lett. 285 (1998) 15; H. Aizawa, Y. Morikawa, S. Tsuneyuki, K. Fukutani, T. Ohno, Surf. Sci. 514 (2002) 394.

[ix] R.M.J. Palmer, A.G. Ferrige, S. Moncada, Nature 327 (1987) 526.

[x] Ze-Haw Zang, Zi-Lin Wu and Shueh-Lin Yau. J. Phys. Chem. B 103 (1999) 9624

[xi] K. Momoi, M-B Song, M. Ito. J. Electroanal. Chem. 473 (1999), 43.

[xii] E. Casero, C. Alonso, J.A. Martín-Gago, F. Borgatti, R. Felici, F. Renner, T.L. Lee, J. Zegenhagen. *Surf. Sci.* 507, 688 (2002).

[xiii] E. Casero, C. Alonso, J.A. Martín-Gago. Electroanalysis 15, Nª8 (2003) 724.

[xiv] J. Clavilier, J. Electroanal. Chem. 107 (1980) 211.

[xv] S Nannarone, F Borgatti, A De Luisa, B P Doyle, G C Gazzadi, A Giglia, P Finetti, N Mahne, L Pasquali, M Pedio, AIP Conference Proceedings 705, 450 (2004); http://www.tasc-infm.it/research/bear/scheda.php

[xvi] M. Benfatto, and S. Della Longa, , J. Synchrotron Rad. 8, 1087 (2001); S. Della Longa, A. Arcovito, M. Girasole, J. L. Hazemann, and M. Benfatto, Phys. Rev. Lett. 87, 155501 (2001)

[xvii] for further detail see M. Pedio, Z. Y. Wu, M. Benfatto, A. Mascaraque, E. Michel, C. Ottaviani, C. Crotti, M. Peloi, M. Zacchigna, and C. Comicioli, Phys. Rev. B **66**, 144109 (2002).

[xviii] J. Stöhr, NEXAFS Spectroscopy, Springer, New York (1992).

[xix] M. Pedio, Z. Y. Wu, M. Benfatto, C. Ottaviani, A. Mascaraque, E. Michel, C. Crotti, M. Peloi, C. Comicioli, Phys. Rev. B 66, pp. 144109-144113, (2002).

[xx] S. Cao, J.-C. Tang, P. Zhu, L. Wang, S. L. Shen, Phys. Rev. B 66 045403 (2002).

[xxi] M. Pedio, M. Benfatto, S. Aminpirooz, J. Haase, Phys. Rev. B **50**, 6596 (1994).

[xxii] H. Geisler, G. Odoerfer, G. Illing, R. Jaeger, H.-J. Freund, G. Watson, E. W. Plummer, M. Neuber, M. Neumann, Surf. Sci. 234, 237-250 (1990).

[xxiii] F. Esch, Th. Greber, S. Kennou, A. Siokou, S. Ladas, R. Imbihl Catalysis Letters 38, 165-170 (1996).

[xxiv] W. Huang, Z. Jiang, J. Jiao, D. Tan, R. Zhai, X. Bao, Surf. Sci. 506, L287 (2002)

[xxv] O. Björneholm, A. Nilsson, H. Tillborg, P. Bennich, A. Sandell, B. Hernnäs, C. Puglia, and N. Mårtensson, Surf. Sci 315 L983 (1994)

[xxvi] F. Sette, Stöhr, E. B. Kollin, D. J. Dwyer, J. L. Gland, J. L. Robbins, A. L. Johson, Phys. Rev. Lett. 54, 935 (1985).

[siesta] J.M. Soler, E. Artacho, J.D. Gale, A. García, J. Junquera, P. Ordejdón, D. Sánchez-Portal, J. Phys.: Condens. Matter **14** 2745 (2002).

[lda] D.M. Ceperley and B.J. Alder, Phys. Rev. Lett. **45**, 566 (1980).

[pseudo] N. Troullier and J.L. Martins, Phys. Rev. B **43**, 1993 (1991).